\documentstyle[12pt]{article}
\input epsf
\newcommand{\be}{\begin{eqnarray}}   
\newcommand{\ee}{\end{eqnarray}}

\newcommand{\gsim}{\lower.7ex\hbox{$
\;\stackrel{\textstyle>}{\sim}\;$}}
\newcommand{\lsim}{\lower.7ex\hbox{$
\;\stackrel{\textstyle<}{\sim}\;$}}

\input epsf

\begin{document}
\begin{titlepage}

\begin{flushright}
SUBATECH-01-07

\end{flushright}
\vspace*{3cm}

\begin{center}
{\Large \bf    Tenacious Domain Walls   in Supersymmetric $QCD$ }
\vspace{2cm}

{\Large  A.V. Smilga} \\

\vspace{0.8cm}

{\it SUBATECH, Universit\'e de
Nantes,  4 rue Alfred Kastler, BP 20722, Nantes  44307, France. }\\

\end{center}

\vspace*{2cm}

\begin{abstract}
We study the structure of the tenacious (existing for all values of masses
of the matter fields) BPS domain walls interpolating between
different chirally asymmetric vacua in supersymmetric QCD  
 in the limit of large masses.
We show that the wall consists in this case of three layers: two outer
layers form a  ``coat''
with the characteristic size $\sim \Lambda^{-1}_{\rm SYM}$ and there is also
the core 
with  width $\sim m^{-1}$. The core always carries a significant fraction
of the total wall energy. This fraction depends on $N_f$ and on the
``windings'' of the matter fields. 

\end{abstract}

\end{titlepage}

\section{Introduction}

The dynamics of supersymmetric gauge theories 
with or without additional matter multiplets
 attracted the attention of theorists since the
beginning of the eighties. It is very well known  \cite{brmog} 
that the pure supersymmetric Yang--Mills theory (SYM), as well as
a class of theories involving extra  matter supermultiplets (SQCD),
based on the $SU(N_c)$ gauge group, involve $N_c$ different
chirally asymmetric vacuum states characterized by the 
different phases of the gluino condensate
  \be
  \langle {\rm Tr}\ \lambda^2 \rangle  \ =\ \Sigma e^{2\pi i k/N_c},\ \ \
  \ \ \  k = 0, \ldots, N_c-1 \ .
  \label{cond}
   \ee
   It was argued recently \cite{Kovner} that on top of $N_c$ chirally
asymmetric vacua (\ref{cond}), also a chirally symmetric vacuum with
zero value of the condensate exists.

The presence of different degenerate physical vacua in the theory
implies the existence of domain walls --- static field configurations 
depending  only on one spatial coordinate ($z$) which interpolate between
one of the vacua at $ z = -\infty$ and another one at $z = \infty$ and
minimizing the energy functional. As was shown in \cite{Dvali,my}, in many
cases the energy density of these walls can be found exactly due to the
fact that the walls present the BPS--saturated states :
  \be
   \label{eps}
   \epsilon \ =\ \frac {N_c}{8\pi^2} \left| \langle {\rm Tr}\ \lambda^2 
\rangle_\infty
    \ -\ \langle {\rm Tr}\ \lambda^2 \rangle_{-\infty} \right| \ ,
    \ee
where the subscript $\pm \infty$ marks the values of the gluino
condensate at spatial infinities. The right side of Eq.(\ref{eps}) 
presents an 
absolute   lower bound for the energy of any field configuration interpolating
between different vacua.
    
    The relation (\ref{eps}) is valid {\it assuming} that the wall is
BPS--saturated. However, whether such a BPS--saturated domain wall
exists or not is a non--trivial dynamic question which can be answered
only in a specific study of a particular theory in interest.

In Refs. \cite{mysVes,jaBPSN} this question was 
studied in the theories involving
$N_f = N_c - 1$ different quark and squark flavors. 
(Each flavor corresponds to a pair of chiral supermultiplets 
$S_f$ and $S_f'$ with opposite chiralities)
These theories are distinguished by the fact that the vacuum expectation
values of squark fields give the mass to {\it all} gauge bosons of the group
$SU(N_c)$ due to Higgs mechanism. Also, when the mass $m$ of the matter
fields is small $m \ll  \Lambda_{\rm SQCD}$, the effective coupling constant
is small and the dynamics of the theory can be analyzed perturbatively.

In particular, the low energy dynamics of the theory in the Higgs phase is
described by the Affleck--Dine--Seiberg effective lagrangian  for the
composite chiral superfields ${\cal M}_{ij} = 2S'_iS_j$ \cite{brmog}. 
It has the Wess--Zumino nature with the superpotential
  \be
\label{ADS}
 W \ = \ - \frac{2(N_c - N_f)}{3 ({\rm det} {\cal M})^{1/(N_c - N_f)}}
- \frac m2 {\rm Tr} {\cal M}\ .
  \ee
When writing Eq. (\ref{ADS}), we assumed that all quark/squark flavors are
endowed with the same small mass $m$. For future purposes, 
we have left $N_f$ as
a free parameter (with the restriction $N_f < N_c$). From now on we set $\Lambda_{\rm SQCD} = 1$. 

 It is not difficult to see that the corresponding 
potential  for the scalar components  ${\mu}_{ij}$ of ${\cal M}_{ij}$,
  \be
 \label{Umij}
U({\mu}_{ij}, \bar{\mu}_{ij}) \ =\ 
\sum_{ij} \left|   \frac{\partial W}{\partial {\mu}_{ij}} \right|^2\ ,
  \ee
 has $N_c$ 
degenerate supersymmetric minima. The chiral condensate at the vacua is given
by the relation
  \be
\label{SigW}
\langle {\rm Tr}\ \lambda^2 \rangle_{\rm vac} \ = \ 
\frac {16\pi^2}{N_c}  W({\rm vac}) \ .
 \ee
It has the form (\ref{cond}) with 
  \be
\label{SigNf}
 \Sigma \ = \  \frac {32\pi^2}{3} \left( \frac {3m}{4N_f} \right)^{N_f/N_c}\ .
  \ee
The ADS effective lagrangian has a Wilsonian nature in a sense that the
characteristic mass of the Higgs field excitations it describes is of order
$m$, which is much
smaller than the mass of the heavy gauge bosons. It is important to understand,
however, that it was derived under the {\it assumption} 
that the relevant values of $|{\mu}_{ij}|$ are large. This is not true near
the chirally symmetric vacuum, where $\langle  {\mu}_{ij} \rangle = 0$, 
and there is no wonder that the latter is not seen in the ADS effective 
lagrangian framework (see \cite{my} for detailed discussion).

Adopting the simplest ansatz 
 \be
\label{flavsym}
{\cal M}_{ij} = \delta_{ij}X^2 \ \ {\rm and\  hence} \ \ \mu_{ij} = 
\delta_{ij} \chi^2
 \ee  
 and adding to the potential
(\ref{Umij}) the  kinetic term $|\partial_\mu \chi|^2$, we can obtain
the BPS wall solutions which interpolate between different chirally asymmetric
vacua. For the theory with $N_c = 2, N_f = 1$, an analytic solution exists 
 \cite{my} . In other cases the solutions can be found numerically 
(see Ref. \cite{jaBPSN} and Sect. 3 of this paper). 

When mass $m$ is not small, the lightest states in the spectrum have the 
glueball/glueballino nature and one cannot write down a truly Wilsonian 
effective lagrangian. However, the situation is better here than, say, in pure
nonsupersymmetric Yang--Mills theory. In our case, the {\it
potential} part of the effective lagrangian is rigidly fixed by symmetry 
considerations. It is expressed in terms of ${\cal M}_{ij}$ and of the 
colorless chiral superfield 
  \be
 \label{Sdef}
 S \equiv \Phi^3 \ =\ \frac 3{32\pi^2} {\rm Tr} \{W_\alpha W^\alpha \}
  \ee
representing the gauge sector. The lowest component of $S$ is proprotional to
$ {\rm Tr}\ \lambda^2 $. Supersymmetry and the exact relations for the 
conformal and chiral anomalies dictate the following form of the 
superpotential \cite{VY}
  \be
\label{WTVY}
 W \ =\ \frac 23 \Phi^3  \left[ \ln \left( \Phi^{3(N_c - N_f)} \det {\cal M}
\right)
 - (N_c - N_f) \right] - \frac m2 \ {\rm Tr} \ {\cal M}  
  \ee
However, the kinetic term of the lagrangian is not fixed rigidly, even though
the requirement of the absence of extra dimensionfull parameter imposes
significant restrictions. The simplest choice is
  \be
  \label{Kahler}
{\cal L}^{\rm kin} \ =\ \int d^4\theta \left[\bar \Phi \Phi + 
{\cal K} (\bar  {\cal M},  {\cal M} ) \right] \ ,
  \ee
where the K\"ahler potential ${\cal K} (\bar  {\cal M},  {\cal M} )$
is the same as in the ADS lagrangian. It
is obtained from the  term $\sum_i (\bar S_i S_i +
\bar S'_i S'_i)$ in the original SQCD lagrangian, which describes physics
adequately for large values of moduli. The sum of Eq. (\ref{Kahler})
and Re $[\int d^2\theta W(\Phi, {\cal M}_{ij})] $ is called
{\it Taylor--Veneziano--Yankielowicz} (TVY) effective lagrangian.

Domain walls in supersymmetric QCD with $N_f = N_c - 1$ were studied in the 
TVY framework in Refs.\cite{my}-\cite{jaBPSN}. The results are the following:
 \begin{enumerate}
\item On top of the chirally asymmetric vacua (\ref{cond}), the system
also enjoys the chirally symmetric vacuum with $\langle \phi^3 \rangle = 
\langle \mu_{ij} \rangle = 0$. 
 \item For any value of mass there are  ``real'' (i.e. without essential
complex dynamics) BPS solutions interpolating between
the chirally symmetric vacuum and each chirally asymmetric one. 
 \item For small masses there are also {\it two different} complex 
BPS wall solutions   interpolating between adjacent chirally asymmetric
vacua. In the limit $m \to 0$, one of these solutions (the ``upper BPS
branch'') goes over to the BPS solution in the ADS effective lagrangian.
Another solution (the lower BPS branch) passes near the chirally
 symmetric minimum and is not described by the ADS lagrangian.
 \item When mass grows, two BPS branches approach each other. They fuse
at some critical value $m_*$. For $m > m_*$, there is no BPS solution at all.
A  domain wall still exists in the range $m_* < m < m_{**}$, but it is 
no longer BPS saturated. At  $m > m_{**}$, there are no such walls 
whatsoever. We have studied the theories with $N_c = 2,3,4$. The analysis
 was later
extended to larger $N_c$ (up to $N_c = 8$) \cite{Binosi}. The critical
value $m_*$ falls off rapidly with $N_c$, while  $m_{**}$ is roughly
constant. 
 \end{enumerate}

In recent \cite{Moreno},  theories with arbitrary number of flavors were
analyzed along the same lines. It was found that, at $N_f < N_c/2$, we
have a completely different picture. Namely, there is only one 
complex BPS branch and it exists for {\it any} value of mass. This finding
was confirmed in \cite{Binosi}. A qualitative explanation of this phenomenon
was given in \cite{Moreno}. In particular, the limit $m \to \infty$
was explored. It was noted that the profile of the wall acquires for
large masses a universal form, which can be found in the framework of the
VY effective lagrangian for pure supersymmetric Yang--Mills theory.

 In addition, it was noticed  that such ``tenacious'' domain walls  also 
exist in 
the theories with $N_f = N_c-1$, if one relaxes the requirement (\ref{flavsym}).
Assuming that ${\cal M}_{ij}$ is still diagonal, but its different components
are not equal, one is able to construct the complex domain walls that 
persists for arbitrary large masses.

In this paper we are making two remarks. We note that, when $m$ is large, 
``tenacious'' walls have a complex ``matryoshka'' structure. 
It involves the VY ``coat'', for which heavy
matter fields decouple, and the core, where the moduli ${\mu}_{ij}$
are  ``alive''. The core has small width $\sim 1/m$, but as the fields  
 ${\mu}_{ij}$ 
change rapidly there and the energy density is big, 
it carries a significant fraction of the total 
wall energy, which we calculate. 

Another remark is that the flavor asymmetric walls found in the second
paper in Ref.\cite{Moreno} exist also at small masses and can be described
in the framework of the ADS effective lagrangian. We present the simplest
such asymmetric solution.

In the last section, we discuss the relevance of these new findings for 
the dynamics of pure supersymmetric YM theory, including the toron 
controversy.          

\section{Tenacious walls at large masses.}

Let us consider first flavor--symmetric walls with the ansatz 
(\ref{flavsym}) for the moduli  ${\cal M}_{ij}$. The superpotential 
(\ref{WTVY}) acquires the form 
  \be
\label{WFiX}
 W \ =\ \frac 23 \Phi^3  \left[ \ln \left( \Phi^{3(N_c - N_f)} X^{2N_f}
\right)
 - (N_c - N_f) \right] - \frac {mN_f}2 X^2\ .   
  \ee
The corresponding scalar field potential is 
  \be
\label{Ufihi}
U(\phi, \chi) \ =\ \left| \frac {\partial W}{\partial \phi} \right|^2 
+ \left| \frac {\partial W}{\partial \chi} \right|^2 \ =\ \nonumber \\
4 \left| \phi^2 \ln (\phi^{3(N_c - N_f)} \chi^{2N_f} )\right|^2  
+ N_f^2 \left| \frac {4\phi^3}{3\chi} - m\chi \right|^2 \ .
  \ee
We are set to study the wall which interpolates between the vacua:
  \be
\phi^3 : \ \ R_*^3 \longrightarrow  R_*^3  e^{2\pi i N_f/N_c}; \ \ \ \ 
\chi^2 : \ \ \rho_*^2 \longrightarrow  \rho_*^2 e^{2\pi i N_f/N_c}
  \ee
with 
\be
\label{Rro}
R_*^3 = \left( \frac {3m}4 \right)^{N_f/N_c}, \ \ \ \ 
\rho_*^2 = \left( \frac {3m}4 \right)^{ N_f/N_c - 1} \ .
  \ee
The BPS equations for the wall have the form 
  \be
 \label{BPSTVY}
\partial_z \phi \ =\ e^{i\delta} 2\bar \phi^2 
 \ln (\bar \phi^{3(N_c - N_f)} \bar \chi^{2N_f} )\ ,
\ \ \ \ \ \partial_z \chi \ =\ e^{i\delta} N_f\left[ 
\frac {4\bar \phi^3}{3\bar \chi} - m\bar \chi \right] 
 \ee
with $\delta = \pi N_f/N_c - \pi/2$. There is an integral of motion
  \be
\label{Wintmot}
{\rm Im} \left[We^{-i\delta} \right] = {\rm Re} 
\left[We^{-i\pi N_f/N_c} \right] \ =\ {\rm const} \ .
  \ee
 The energy of the wall is 
 \be
\label{Ewall}
 \epsilon = 2|W_\infty - W_{-\infty}| \ =\ 
\frac {4N_c R_*^3}3 \left|    e^{2\pi i N_f/N_c} - 1 \right|
= \frac {8N_c R_*^3}3 \sin \frac {\pi N_f} {N_c} \ .
 \ee
Note that the phase of the fields $\phi$ and $\chi$ must change along
the wall in such a way that the phase of the argument 
of the logarithm in Eq. (\ref{Ufihi}) remains zero at $z = -\infty$
as it is at $z = \infty$ :
  \be
\label{Delarg}
(N_c - N_f) \Delta \ {\rm arg} [\phi^3 ] + N_f 
\Delta \ {\rm arg} [\chi^2 ] \ =\ 0\ .
  \ee
For large masses the matter field $\chi$ tends to get frozen in such
a way that the potentiallly large second term in Eq.(\ref{Ufihi})
vanishes: 
  \be
\label{freeze}
\chi^2 = 4\phi^3/(3m)\ .
  \ee
 The effective potential for the light
field $\phi$ acquires the VY form :
  \be
 \label{UVY}
U^{VY}(\phi) \ =\ 4N_c^2 \left|
\phi^2 \ln (\phi^3/R_*^3) \right|^2 \ .
 \ee
The potential (\ref{UVY}), as it is written, has only one minimum at
$\phi^3 = R_*^3$ and not $N_c$ minima as we expect it to have. The resolution
of this apparent paradox is well known \cite{Kovner} : one should take
different branches of the logarithm at different values of $\phi^3$.
The branches are glued together
\footnote{ Glued potentials are not specific for supersymmetric
theories and also appear  in the Schwinger model \cite{QCD2}.} at 
$$\phi^3 = R_*^3 \exp \{i\pi (1 + 2k)/N_c \} \ , \ \ \ \ \ \ \ 
k = 0,\ldots,N_c-1 \ .$$
Each branch has its own vacuum with $\langle \phi^3 \rangle_k  = 
R_*^3 e^{2\pi i k/N_c}$. Actually, one can see how the branches and the
branch cuts appear in the framework of the TVY model. The point is that the
condition (\ref{freeze}) cannot be satisfied {\it everywhere}: it would
contradict the requirement (\ref{Delarg}). The only way for the solution
to satisfy the both contradicting requirements is the following: the relation
(\ref{freeze}) holds almost everywhere in the wall but for the narrow
central region, where the field $\chi$ changes rapidly such that 
 \be
\label{Delargcor}
\Delta_{\rm core} \ {\rm arg} [\chi^2] \ =\ -2\pi \ .
 \ee
As a result of such a change, the argument of the logarithm in the 
effective theory (\ref{UVY}) 
is multiplied by $ e^{-2\pi i N_f/N_c}$ and this exactly corresponds
to crossing the branch cut and going over to another branch of the glued
potential. 

This scenario works, indeed, in many cases. It is clearly seen from
the numerical solutions of Refs.\cite{Moreno,Binosi}. 
Take Figs. 1,2 of Ref.\cite{Moreno}a. One can see that the variable
$R = |\phi|$
just follows the solution of the effective VY theory. The variable
$\rho = |\chi|$ is frozen according to Eq. (\ref{freeze}) everywhere
but in the central region, where it undergoes a  rapid change. The same
concerns the phase $\beta$ of the variable $\phi$ vs. the phase $\alpha$
of the variable $\chi$.   

\begin{figure}
   \begin{center}
        \epsfxsize=300pt
        \epsfysize=150pt
        \vspace{-5mm}
        \parbox{\epsfxsize}{\epsffile{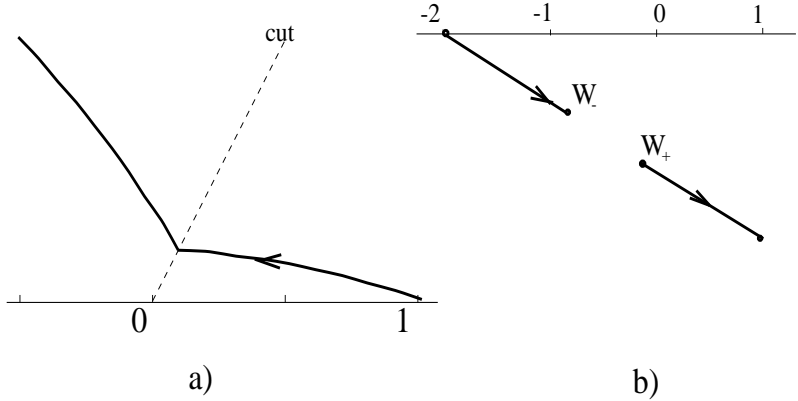}}
        \vspace{-5mm}
    \end{center}
\caption{Argand plots for {\it a)} $\phi^3/R_*^3$ and 
{\it b)} $W(\phi^3)/R_*^3$ for the
BPS domain walls in the effective VY theory; $N_c = 3$. $W_\pm$ are the values
of the superpotential on the opposite sides of the cut.}
\label{figVY}
\end{figure}

To acquire further understanding, we plot in Fig.\,\ref{figVY} the Argand plots 
for $\phi(z)$ and for 
the  superpotential $W[\phi(z)]$ in VY theory. $W$ changes along a straight
line due to the property (\ref{Wintmot}). We see that 
the  superpotential is discontinuous on the cut.

 The value of $\phi^3$ on the cut is given by 
  \be
 \label{phicentr}
(\phi^3)_0 \ =\ R_*^3 \eta e^{i\pi N_f/N_c}\ ,
 \ee
where $\eta$ satisfies the condition
 \be
\label{condeta}
\eta (\ln \eta - 1) \ =\ - \cos \frac {\pi N_f}{N_c}\ ,
  \ee  
which is a corollary of Eq.(\ref{Wintmot}). 

\begin{figure}
   \begin{center}
        \epsfxsize=300pt
        \epsfysize=150pt
        \vspace{-5mm}
        \parbox{\epsfxsize}{\epsffile{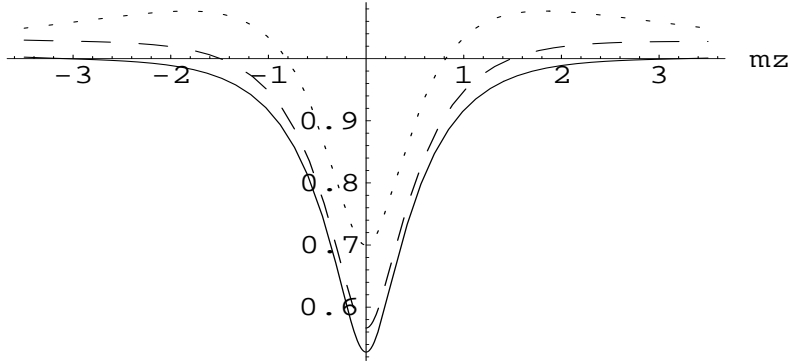}}
        \vspace{-5mm}
    \end{center}
\caption{Dynamics of the the field $\rho_\zeta(z)$ inside the core: $N_c = 3,
m = 50$ (dotted), $N_c = 3, m = 250$ (dashed), $m \to \infty$ (solid).}
\label{figzet}
\end{figure}

Now, in TVY theory with large but finite $m$ there is no discontinuity, but
a narrow transitional region. Within the core one can assume that $\phi^3$
is given by Eq.(\ref{phicentr}) and is constant. It is convenient to introduce
  \be
 \label{zeta}
\zeta \ =\ \sqrt{\frac {3m}{4\phi^3}} \chi \ .
 \ee
The equation describing the dynamics of $\zeta$ in the core has the 
universal form
  \be
\label{BPSzeta}
 \partial_z \zeta \ = \ -im\left( \frac 1{\bar \zeta} - \bar \zeta \right) \ .
 \ee
The solution to Eq.(\ref{BPSzeta}) with boundary conditions $\zeta(\pm \infty)
 = 1$ can be easily found with Mathematica. [see  Fig. \ref{figzet}, where
it is plotted together with the right side of Eq.(\ref{zeta}) obtained
  from the numerical solutions of the BPS equations in TVY
theory for large but finite $m$.]
The phase $\gamma$  ($\zeta \equiv \rho_\zeta e^{i\gamma} $) is changed by 
$-\pi$. There is an integral of motion
 \be
\label{zintmot}
 \ln \rho_\zeta - \frac {\rho_\zeta^2}2 \cos (2\gamma) 
\ =\ {\rm const} = - \frac 12\ .
 \ee
In the center of the wall, $\gamma = -\pi/2$ and $\rho_\zeta \approx 
0.52$. We see from Fig \ref{figzet} that, for large masses, 
 the dependence $|\chi(z)|$ 
inside the core is determined by Eq.(\ref{BPSzeta}), indeed. 
 Also, $\chi(z)$  satisfies 
the condition (\ref{freeze}) in the coat.

\begin{figure}
   \begin{center}
        \epsfxsize=300pt
        \epsfysize=150pt
        \vspace{-5mm}
        \parbox{\epsfxsize}{\epsffile{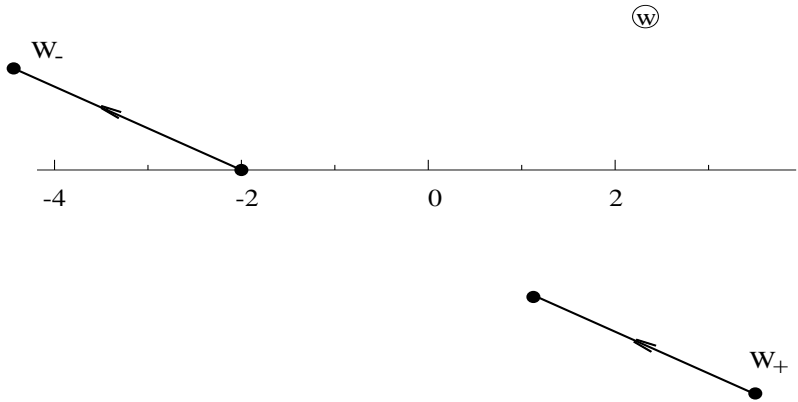}}
        \vspace{-5mm}
    \end{center}
\caption{The coat for a nonexisting wall ($N_c = 3$).}
\label{figeta2}
\end{figure}

Let us determine the fraction of the energy of the wall stored in its coat.
It is given by the expression 
  \be
 \label{dolaW}
f_{\rm coat} \ =\ \frac {|W_\infty - W_+| + |W_- - W_{-\infty}|}
{|W_\infty - W_{-\infty}|} \ ,
 \ee
where $W_\pm$ are the values of the superpotential at the opposite sides of 
the core. A simple calculation using, again, the condition (\ref{Wintmot})
gives 
 \be
 \label{dola}
f_{\rm coat} \ =\ \frac {\left| \sin  \frac {\pi N_f}{N_c}
-   \frac {\pi \eta N_f}{N_c}\right|}{\sin  \frac {\pi N_f}{N_c}} \ .
 \ee
Let us look at Eq.(\ref{condeta}) determining the parameter $\eta$.
At $N_f/N_c < 1/2$ it has two real roots. One of them ($\eta_1$) 
is smaller than 1 and the corresponding fraction (\ref{dola}) is also
less than 1. Another root ($\eta_2$) lies within the range
$1 < \eta_2 < e$. For $N_c < 5$,  the corresponding fraction is greater 
than 1, which obviously 
means that this solution is not acceptable. But even for
$N_c > 5$ when $f_{\rm coat}$ as determined by Eq.(\ref{dola}) is less
than 1, the root $\eta_2$ does not correspond to any wall solution
in TVY theory. Again, the picture can be 
clarified by drawing the Argand plot for the corresponding BPS solutions
in VY theory (see Fig. \ref{figeta2}). The differences
$W_\infty - W_+,\ W_- - W_{-\infty}$ have now the opposite sign as 
compared to the previous case, and one can no longer 
pass from  $W_-$ to $W_+$ moving in the positive $z$ direction. 

If $N_f/N_c = 1/2$, $\eta_1 = 0$. This corresponds to the wall passing
through the chirally symmetric vacuum in the middle. 
This is, indeed, the only way for different chirally asymmetric vacua 
to be connected in the theory with $N_c = 2, N_f = 1$ for large masses
\cite{mysVes}. For  $N_f/N_c > 1/2$, the real root $\eta_1$ disappears
and there is no solution whatsoever. 

If $N_f = 1$ and $N_c$ is large, $\eta \approx 1 - \pi/N_c$ and 
  \be
 \label{largeN}  
 f_{\rm coat}^{{\rm 1\ flavor, \ large }\ N_c} \ \approx
\ \frac {\pi}{N_c} \ll 1 \ ,
 \ee
i.e. almost all energy is stored in the core. This agrees with the analysis
in Ref.\cite{KKS}, where the mechanism of regularizing the branch cut
singularity by ``integrating in'' an extra heavy field was first suggested.
(The authors of  \cite{KKS} did not analyze  TVY theory, however, and
restricted themselves to discussion of toy models.) If not only $N_c$, but
also $N_f$ is large, the arguments of \cite{KKS} do not apply and 
$ f_{\rm coat}$ is not necessarily small. The argument based on the analysis 
of the expression (\ref{dola}) gives an explanation why flavor--symmetric
walls do not exist when  $N_f/N_c > 1/2$ and $m$ is large, which is 
complementary to that in Refs.\cite{Moreno}. 

It was noticed that, if the
requirement (\ref{flavsym}) is relaxed, tenacious domain walls exist even
in the range   $N_f/N_c > 1/2$. Consider the simplest case $N_c = 3,\ N_f = 2$.
Assume ${\mu}_{ij} = {\rm diag}(\chi_1^2, \chi^2_2)$, 
$\chi_1 \neq \chi_2$. A tenacious BPS solution with
 \be
\label{Delasym}
\Delta \ {\rm arg} [\phi^3]  =   \Delta \ {\rm arg} [\chi_1^2] \ =\ 
\frac {2\pi}3, \ \ \ \ \  
 \Delta \ {\rm arg} [\chi_2^2]    = -\frac {4\pi}3 
  \ee
exists. 
In the large mass limit $\chi_1^2$ stays frozen according to Eq.(\ref{freeze})
{\it everywhere}, while the field $\chi_2^2$ undegoes a rapid change in the
core, which is described by the universal equation (\ref{BPSzeta}). Using the
terminology of Ref.\cite{Moreno}b, the field $\chi_2$ has a notrivial 
{\it winding}, while $\chi_1$ has not. As $\Delta \ {\rm arg} [\phi^3]$
is the same in this case as in the theory with $N_c = 3,\ N_f = 1$, the Argand
plots for $\phi^3$ and $W(\phi^3)$ in the effective VY theory 
are also the same and are given in Fig. \ref{figVY}. The wall exists  for 
arbitrary large masses.
The fraction of the energy 
stored in the coat   is given by the same formula (\ref{dola}),
with $N_f$ substituted by 
  \be
\label{k}
k = \sum_i^{N_f} \omega_i\ ,
 \ee
  where $\omega_i$
are the windings of the matter fields. They acquire values 0 or 1. 
A wall with $k/N_c < 1/2$ is tenacious.

\section{Tenacious walls in the ADS limit.}

The main characteristic feature of the tenacious solutions is that they persist
for arbitrary large masses. But, of course, they also exist in the small
mass limit, where the system is described by the ADS effective
lagrangian with the superpotential (\ref{ADS}). The latter is obtained
from Eq.(\ref{WTVY}) by freezing the heavy field $\Phi$ so that the argument
of the logarithm is equal to 1. 

Note that fractional powers in Eq.(\ref{ADS}) do not give rise to a 
new kind of  glued potentials. The point is that the domain wall solutions
always stay on the same sheet of the function (\ref{ADS}) and the problem
of discontinuities associated with branch cuts does not arise. Suppose
e.g. that $N_f = 1$. Then
 \be
\label{ADSchi}
W \ =\  - \frac {2(N_c - 1)}{3(X^2)^{1/(N_c - 1)}} - \frac m2 X^2 \ .
 \ee
Let us choose the sheet where $W$ is real for real positive $X^2$. Then
a perfectly smooth BPS domain wall interpolating between $\chi^2 = \rho_*^2$
and  $\chi^2 = \rho_*^2 e^{2\pi i/N_c}$ exists such that 
  \be
 \label{DelADS}
\Delta \ {\rm arg} [\chi^2] \ =\ \frac {2\pi}{N_c} - 2\pi ,\ \ \ \ \  
 \Delta \ {\rm arg} \left[(\chi^2)^{-1/(N_c - 1)}\right] \ =\ 
\frac {2\pi}{N_c}\ .
 \ee
One can remind here that, though the theories with $N_f = N_c -1$ are somewhat
nicer because all gauge fields become heavy and we are in the Higgs weak
coupling regime, the mass of the lowest excitations is of order 
$m \ll \Lambda_{SQCD}$ for any $N_f$, and the ADS effective lagrangian has
always a Wilsonian nature. 

The ADS lagrangian does not describe, however, the chirally symmetric sector,
where $\Phi^3 \equiv 0$ and the effective superpotential is just
 \be
 \label{Wchinv}
 W^{\rm chir. inv.}_{\rm eff}({\cal M}) \ =\ 
- \frac m2 {\rm Tr} {\cal M}\ .
 \ee
The walls that penetrate into this sector also have, for small masses, a 
multilayer matryoshka structure similar to the structure of
tenacious walls in the large mass limit, discussed above, only the role
of the heavy and light fields is now reversed. 

This especially
concerns the ``lower BPS branch'' for the flavor--symmetric non--tenacious 
walls. As was shown in Sect. 7 of Ref.\cite{jaBPSN}, the wall in this case
consists of {\it five} layers: 
 \begin{itemize}
\item two wide (with characteristic width $\propto 1/m$) outer layers whose
dynamics is described by the ADS lagrangian,
\item a wide central region, where $\phi^3$ is close to zero and there is only
the quadratic term $\propto |\bar \chi \chi|$ in the effective potential,
\item two narrow (with the characteristic width $\propto \Lambda_{SQCD}$)
transitional regions, where the field $\phi^3$ changes rapidly, while the 
matter field $\chi$ stays effectively frozen.
\end{itemize}

Also ``real'' walls interpolating between the chirally symmetric and 
 chirally asymmetric vacua consist of two layers: a narrow one, where
the field $\phi^3$ changes rapidly and a wide layer, where $\phi^3 \approx
0$ and only the matter field changes.

\begin{figure}
\begin{minipage}[t]{70mm}
   \begin{center}
        \epsfxsize=200pt
        \epsfysize=100pt
        \vspace{-5mm}
        \parbox{\epsfxsize}{\epsffile{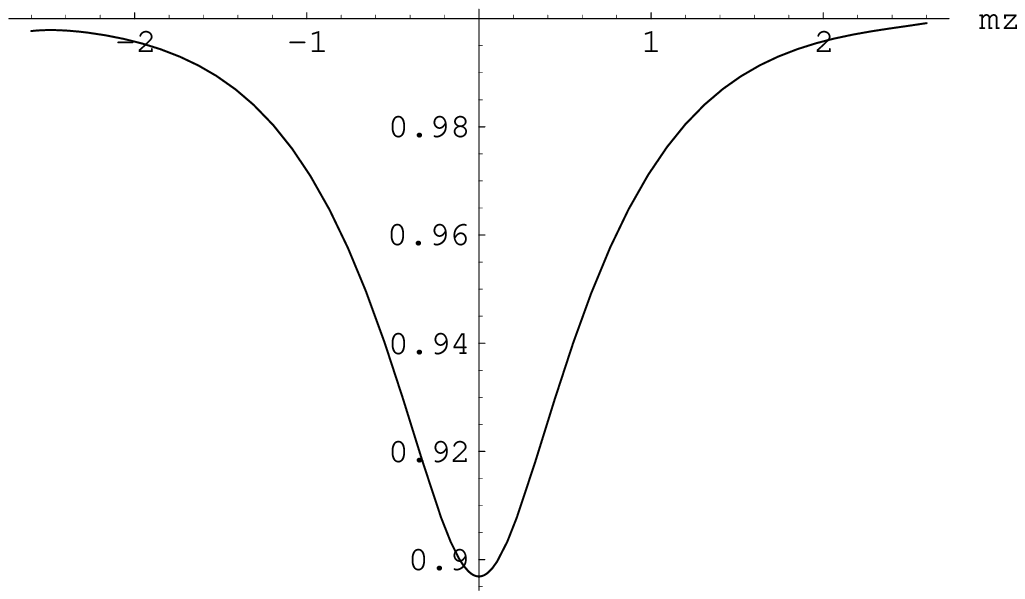}}
    \end{center}
\end{minipage}
\begin{minipage}[t]{70mm}
   \begin{center}
        \epsfxsize=200pt
        \epsfysize=100pt
        \vspace{-5mm}
        \parbox{\epsfxsize}{\epsffile{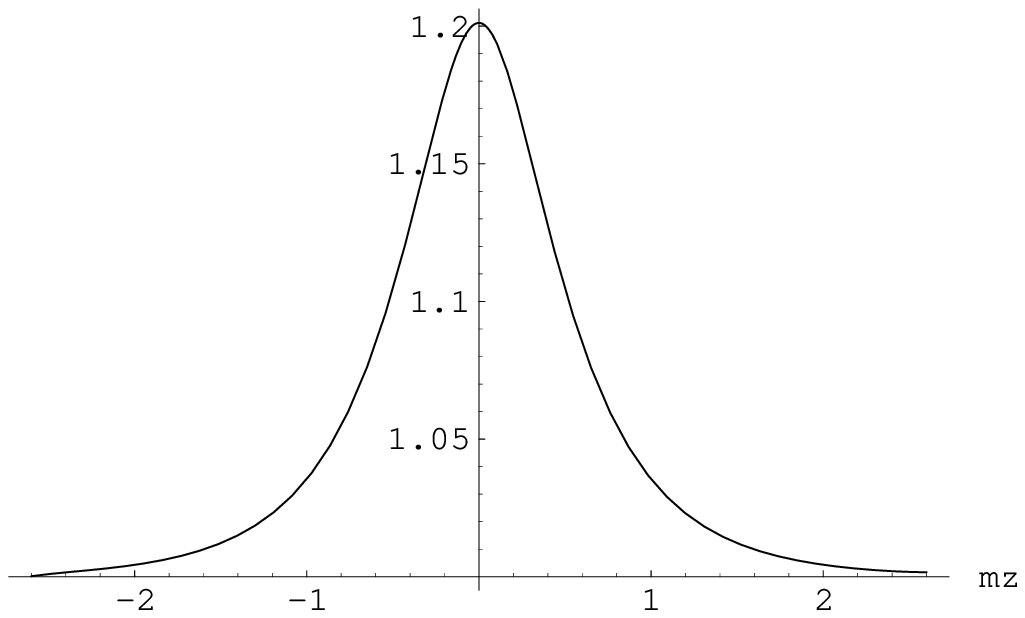}}       
    \end{center}
\end{minipage}
\caption{Asymmetric wall in the BPS limit ($N_c = 3,\ N_f = 2,\ k=1$): 
the plots of $\rho_1(z)/\rho_*$ (left) and $\rho_2(z)/\rho_*$ (right).}
\label{figasym}
\end{figure}

The ADS lagrangian  also admits flavor--asymmetric wall solutions. 
Consider the case $N_c = 3, N_f = 2,\ 
{\cal M}_{ij} = {\rm diag}(X_1^2, X^2_2)$. The superpotential is
   \be
\label{ADS12}
W \ =\  - \frac {2}{3X_1^2X_2^2} - \frac m2 (X_1^2 + X_2^2) \ .
 \ee
Consider the wall where the phases of the fields $\chi_1^2$, $\chi_2^2$
change in opposite directions:
  \be
\chi_1^2 : \  \rho_*^2 \longrightarrow  \rho_*^2  e^{2\pi i /3}; \ \ \ \ \ \ \  
\chi_2^2 : \  \rho_*^2 \longrightarrow  \rho_*^2 e^{-4\pi i/3} \ .
  \ee
The corresponding BPS equations
 \be
 \label{BPS12}
\partial_z \chi_1  &=& e^{-i\pi/6} \left[ \frac 4{3\bar \chi_1^3 \bar \chi_2^2}
- m \bar \chi_1 \right]\ , \nonumber \\
\partial_z \chi_2  &=& e^{-i\pi/6} \left[ \frac 4{3\bar \chi_2^3 \bar \chi_1^2}
- m \bar \chi_2 \right]
  \ee
can be solved. The profiles of $\rho_1 = |\chi_1|$ and $\rho_2 = |\chi_2|$
are presented in Fig. \ref{figasym}. 

The presence of flavor--asymmetric domain walls is a rather remarkable and 
nontrivial fact. Note that their presence in the Wilsonian ADS lagrangian
{\it assures} their existence in SQCD. 
Flavor--symmetric and flavor--asymmetric domain walls exist in the ADS limit
for any $N_f$ and any combination of windings. Lifting them up to TVY theory,
we observe that, for low masses, the argument of the logarithm in 
Eq.(\ref{WTVY}) is close to
1 and its phase --- to zero.   Therefore, in contrast to the authors of
Ref.\cite{Moreno}b, we would not call ``unphysical'' the solutions with
$k > 1$ on the ground that the phase of the logarithm changes by $\Delta
 = 2\pi k > 2\pi$ in the central region of the wall  in the 
large mass
limit. They are certainly physical at small masses when $\Delta = 0$. Then
$\Delta$ increases with mass up to $2\pi k$ for $m \to \infty$, but we do not
see in what respect the solution with $\Delta = 359^o$ at somewhat smaller
mass is better than the solution at somewhat larger
mass with $\Delta = 361^o$. Both solutions are smooth and do not have problems
with discontinuities associated with branch cuts. (One should be very careful,
indeed, when a solution runs into such a discontinuity) 
Unlike the case of flavor--symmetric
walls at $N_f = N_c -1$ analyzed in Refs.\cite{mysVes,jaBPSN}, 
no phase transition in mass occurs here. 

In the case $N_c = 3, N_f = 2, k=1$, we have three walls: two flavor asymmetric walls and the flavor symmetric one. For arbitrary $k$ and $N_c$ ($N_f = N_c - 1$), the number of different walls can be determined by adapting the arguments of Ref.\cite{Vafaold}, where the number of solitons in the supersymmetric $CP^{N-1}$ model was calculated. [The relevant effective Lagrangian for 
$CP^{N-1}$ 
coincides with the ADS Lagrangian in the framework of the diagonal ansatz (\ref{flavsym}).
\footnote{I am indebted to T. van Veldhuis and A. Vainshtein for the discussion of this point.}]
Let us write the superpotential as 
  \be
  \label{CPN}
  W \ =\ - \frac m2 \sum_{i=1}^{N_c}  X_i^2  \ ,
   \ee
   where 
   $$ X_{N_c}^2 \ \stackrel{\rm def}= \ - \frac 4{3m \prod_{i=1}^{N_c - 1} X_i^2}\ . $$
The wall interpolates between the point with the values of the superpotential $W_*$ and $W_* \exp\{2\pi i k/N_c \}$. Now, for each $\chi_i^2$ we have 
  \be
  \label{argchi}
   \Delta  \ {\rm arg} [\chi_i^2] \ =\ 2\pi \left( \frac k{N_c} - \omega_i \right)
 \ee
 with $\omega_i = 0$ or $\omega_i = 1$. 
Bearing in mind that $ \Delta  \ {\rm arg} 
[\prod_{i=1}^{N_c}\chi_i^2] = 0$, there are $k$ fields with 
$\omega_i = 1$ and $N-k$ fields with $\omega_i = 0$. 
Altogether, there are
 $C_k^{N_c}$ possibilities.\footnote{This was done in the framework of the diagonal ansatz (\ref{flavsym}). Allowing for flavor rotations and assuming the mass
matrix to be diagonal, we obtain not a discrete set of walls, but continuous
families of solution (the wall moduli space). It seems probable that the degeneracy of this moduli space is lifted for a generic  mass matrix and we are left with $C_k^{N_c}$ 
isolated solutions. \cite{Ritz}.}
 Remarkably, this coincides with the estimate for the number of walls in SYM theory compactified on $T^2$, obtained in Ref.\cite{Vafanew} using D-brane arguments. We should emphasize again , however, that this counting does not work for the TVY model. In the theory
with $N_f = N_c -1$, there are more than $C_k^{N_c}$ walls in the limit of small masses and less than $C_k^{N_c}$  tenacious walls 
surviving  for large masses. [The  number of the latter is equal to
$C_k^{N_c-1}$ if  $k < N_c/2$ and to zero otherwise --- see Eq.(\ref{k})] \ldots

\section{Discussion}
For us, the main interest of the study of the domain walls in supersymmetric
QCD is  a hope to shed light on the long--standing ``toron controversy'',
associated with the vacuum structure of pure SYM theory. Two different
interpretations of the basic relation (\ref{cond}) are possible:
 \begin{enumerate}
\item The chiral symmetry $U_A(1)$ of the free SYM action is explicitly 
broken down to $Z_{2N_c}$ due to anomaly. Further, $Z_{2N_c}$ is spontaneously
broken down to $Z_2$. The condensate (\ref{cond}) plays the role of order
parameter associated with this breaking. 
\item $U_A(1)$ is explicitly broken down to $Z_2$ and the different
vacua (\ref{cond}) lie in different sectors of the Hilbert space. 
The integer $k$ plays in this case the same role as the parameter $\theta$
or, better to say, $\theta$ changes within the range $(0, 2\pi N_c)$ and
not within the range  $(0, 2\pi) $ as it does in standard QCD.
This implies the 
relevance of ``torons'' --- configurations with fractional topological
charges which exist in a finite 4--dimensional box \cite{Hooft} and
might stay relevant also in the limit when the size of the box
is sent to infinity \cite{toron}. All the arguments {\it pro} and
{\it contra} were discussed recently anew in  Ref.\cite{my}. 
 \end{enumerate}
The first picture (complemented with the assumption that the chirally
asymmetric vacuum is an artifact of the VY approach and {\it is} not really there) 
is standard.\footnote{One of the  arguments is the absence of the chirally symmetric state and irrelevance of torons in  ${\cal N} = 2$
SYM theory \cite{Arc}.} This necessarily implies the presence of physical domain walls interpolating between the vacua with different phases of $\langle {\rm Tr} \{\lambda^2\} \rangle$. If the second picture is correct,  there are no such domain walls.

 At the moment, one cannot say with certainty whether these domain walls exist or not in pure SYM theory. There are D-brane arguments in favor of their existence \cite{Witten}, but it is important to try to resolve this field theory issue within the field theory framework. In early works 
\cite{mysVes,jaBPSN}, it was shown that a certain type of walls present in
supersymmetric QCD disappears in the large mass limit. The results of
Refs.\cite{Moreno, Binosi} and of the present work show that there are
tenacious walls, which persist in the large mass limit. Note, however, that
the core of such a wall becomes very thin in this limit and the energy
density in the core becomes very large. We find this situation rather
queer. The assumption that these walls with narrow dense core are relevant
for physics in pure SYM theory contradicts the common wisdom that heavy fields
should decouple in the limit $m \to \infty$ and have no effect on the 
dynamics of the low energy sector. 

We {\it think} that this general argument should work also in this case, but
obviously, further study of this question is required.

The last comment concerns the dynamics of SYM theory at finite temperature.
Even though supersymmetry is broken by temperature
\footnote{Contrary to what many people think, this breaking is not
explicit, but spontaneous \cite{sound}.}, one can use temperature as a 
theoretic tool to distinguish between two different scenario of the chiral 
symmetry breaking in SYM mentioned above. It is not easy, however.

In the first scenario (with walls), the spontaneously broken discrete
chiral symmetry is restored at some critical temperature $T_c$, by the
same token as the spontaneously broken continuous chiral symmetry $SU_L(N_f)
\times SU_R(N_f)$ is restored in the standard massless QCD. If one assumed
that complex domain walls decouple in the large mass limit {\it and}
the presence of the chirally symmetric vacuum were disregarded, one
would conclude that the chiral condensate retained a nonzero value for any
temperature (as it does in QCD with one light flavor) and there would be
 no phase transition.
 However, in the TVY model different chirally asymmetric vacua {\it can}
communicate with each other with  the chirally symmetric state as an 
intermediary. Practically, this means that the chiral symmetry is duly
restored at some $T_c$ in the same way as it does in the standard scenario.

To conclude with, decoupling of complex walls in the large mass limit implies
{\it either} appearance of a new superselection rule for the parameter $k$
and the relevance  of fractional topological charges in pure SYM theory
{\it or} the presence
of the chirally symmetric vacuum state. The  TVY/ VY approach favors the second
possibility.  

I am indebted to D. Binosi, M. Shifman, A. Vainshtein, and T. van Veldhuis for illuminating discussions.

\end{document}